# Giant Real-time Strain-Induced Anisotropy Field Tuning in Suspended Yttrium Iron Garnet Thin Films

Renyuan Wang[1†], Sudhanshu Tiwari[2*], Yiyang Feng[2], Sen Dai[2], and Sunil A. Bhave[2]

[1]*FAST Labs™, BAE Systems, Inc.*
*65 Spit Brook Road, Nashua, NH 03087, USA*

[2]*Elmore Family School of Electrical and Computer Engineering, Purdue University*
*1205 West State St., West Lafayette, IN 47907, USA*

Yttrium Iron Garnet based tunable magnetostatic wave and spin wave devices are poised to revolutionize the fields of Magnonics, Spintronics, Microwave devices, and quantum information science. The magnetic bias required for operating and tuning these devices is traditionally achieved through large power-hungry electromagnets, which significantly restraints the integration scalability, energy efficiency and individual resonator addressability. While controlling the magnetism of YIG mediated through its magnetostrictive/magnetoelastic interaction would address this constraint and enable novel strain/stress coupled magnetostatic wave (MSW) and spin wave (SW) devices, effective real-time strain-induced magnetism change in YIG remains elusive due to its weak magnetoelastic coupling efficiency and substrate clamping effect. We demonstrate a heterogeneous YIG-on-Si MSW resonator with a suspended thin-film device structure, which allows significant straining of YIG to generate giant magnetism change in YIG. By straining the YIG thin-film in real-time up to 1.06%, we show, for the first time, a 1.837 GHz frequency-strain tuning in MSW/SW resonators, which is equivalent to an effective strain-induced magnetocrystalline anisotropy field of 642 Oe. This is significantly higher than the previous state-of-the-art of 0.27 GHz of strain tuning in YIG. The unprecedented strain tunability of these YIG resonators paves the way for novel energy-efficient integrated on-chip solutions for tunable microwave, photonic, magnonic, and spintronic devices.

Single crystal Yttrium iron garnet (YIG, $Y_3Fe_5O_{12}$) exhibits the lowest magnon damping among known magnetic materials[1]. Fueled by the material's many interesting cross-physical-domain coupling properties, there have been many explorations into using YIG to realize devices such as magneto-optic devices for cryogenic applications[2], non-reciprocal optical and RF/microwave devices[3–5], tunable RF/microwave devices[6–8], hybrid quantum circuits for quantum information processing[9–13], spintronic devices[14–17], and devices for realizing room temperature Bose-Einstein condensates[18]. In most of these applications, the magnetic bias needed for operating and/or tuning of the YIG device was achieved through a bulky and power hungry electromagnet, which significantly restraints the integration and energy-efficiency scalability. Electrical control of the magnetism of YIG through its magnetoelastic interaction, for example, through piezoelectric actuation, would enable beyond CMOS scalable energy-efficient devices for high efficiency magnetoelectric energy harvesting[19], ultra-low power non-volatile memory[20], electrically small magnetoelectric antenna[21], spin-state manipulation in quantum sensing and quantum information processing[9,22–27], and novel spintronic and tunable devices[28–38]. However, effective real-time strain or stress mediated control of magnetism in YIG has remained intractable due to considerably small magneto-elastic coefficients compared to other magnetoelastic materials such as terfenol-D and FeGaB[39–41]. This results in the requirement of a large amount of strain to generate a useful strain-induced magnetocrystalline anisotropy field[42]. This is exacerbated by the fact that high quality YIG thin-films are typically grown on lattice matched crystalline bulk substrates[42,43] that are incompatible with modern microfabrication technologies for etching. Consequently, most reported literature on strain control of YIG devices utilize solid mount YIG, which prevents transduction of large amounts of stress/strain in YIG due to substrate clamping[28,36,41,44]. In addition, intimate heterogeneous integration of YIG with piezoelectric and ferroelectric materials is needed to facilitate electrical control of strain/stress transduction, which has been proven challenging to achieve[45,46]. State-of-the-art piezoelectric materials can only achieve 0.3% ~ 0.6% of maximum unloaded strain[47] and under loaded conditions, this is not sufficient to generate a useful amount of strain (therefore strain-induced magnetocrystalline anisotropy field) in YIG with a solid-mounted device structure.

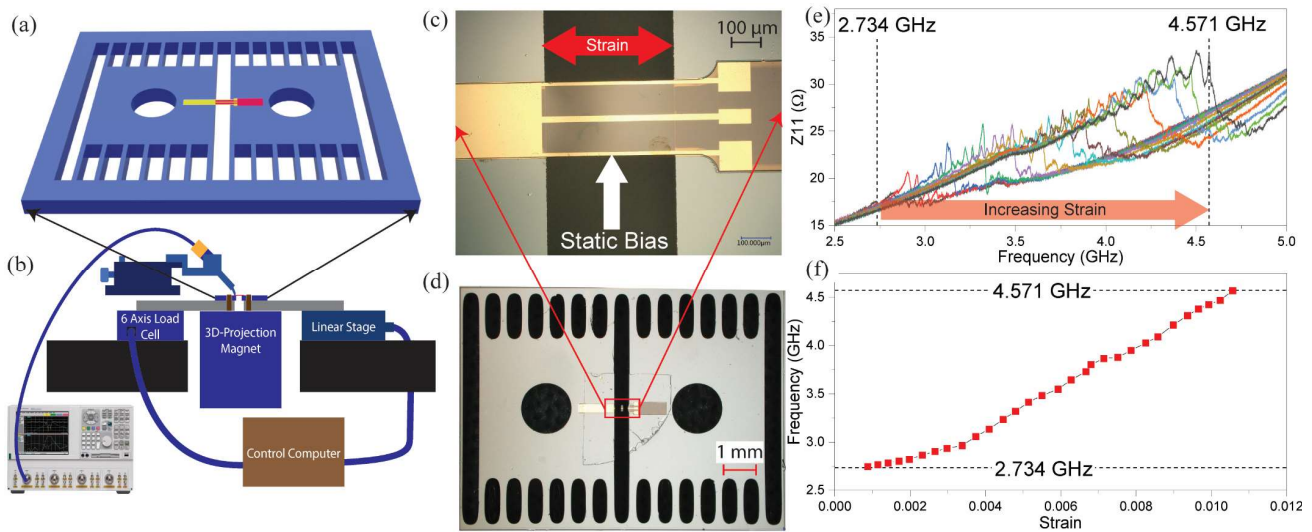

Figure 1: (a) A 3D rendering of the YoS strain tuning device consisting of a YIG MSW resonator suspended over two movable Si shuttles, which are connected to a fixed Si frame by 28 thin Si springs. (b) The measurement setup used for characterization of the strain tunability of the device. (c) An optical image of the suspended YIG resonator and (d) A microscope photo of the strain-tuned resonator device in (a) consisting of a YIG thin-film resonator ion-sliced from a bulk single crystal YIG subsrate, suspended on a bulk micromachined silicon frame. (e) Measured impedance spectra of the device under various levels of applied strain in the YIG resonator. (f) Obtained strain vs frequency response from the measured spectrum as shown in (e).

To address these challenges and enable strong mechanically mediated control of magnetism in YIG, we demonstrate heterogeneous suspended thin-film YIG-on-Si (YoS) magnetostatic wave resonators (Figure 1). The resonators were fabricated on a YoS material platform (Figure 2), with 2.19 µm thick YIG ion-sliced from bulk single crystal substrate, bonded to an oxidized Si wafer using a gold to gold thermocompression bonding process (see Methods). While highly anisotropic etching of YIG is challenging to achieve with conventional wet etching or reactive ion etching techniques, we developed an anisotropic ion mill etching process to pattern thick (>2 µm) YIG thin-film device structures, which is beneficial for minimizing spin-wave scattering. The devices consist of suspended ion-sliced YIG thin film spin-wave resonators anchored on two silicon shuttles connected to the Si substrate through Si spring beams, which are formed by deep reactive ion etching (DRIE) of silicon. The suspended region of YIG has dimensions of 260 µm x 500 µm. Since the YIG is free-standing by removing the substrate, the device structure has significantly lower stiffness compared to the solid mount devices. This allows a large amount of mechanical strain to be transduced in to the free-standing YIG membrane. By actuating the Si shuttles using linear translation actuators, we transduce up to 1.06% of tensile strain in the suspended YIG thin film, which leads to a 1.837GHz of real-time strain induced frequency tuning. This corresponds to a record strain-induced magnetocrystalline anisotropy field of 642 Oe[28,36,41,44,48] (Methods).

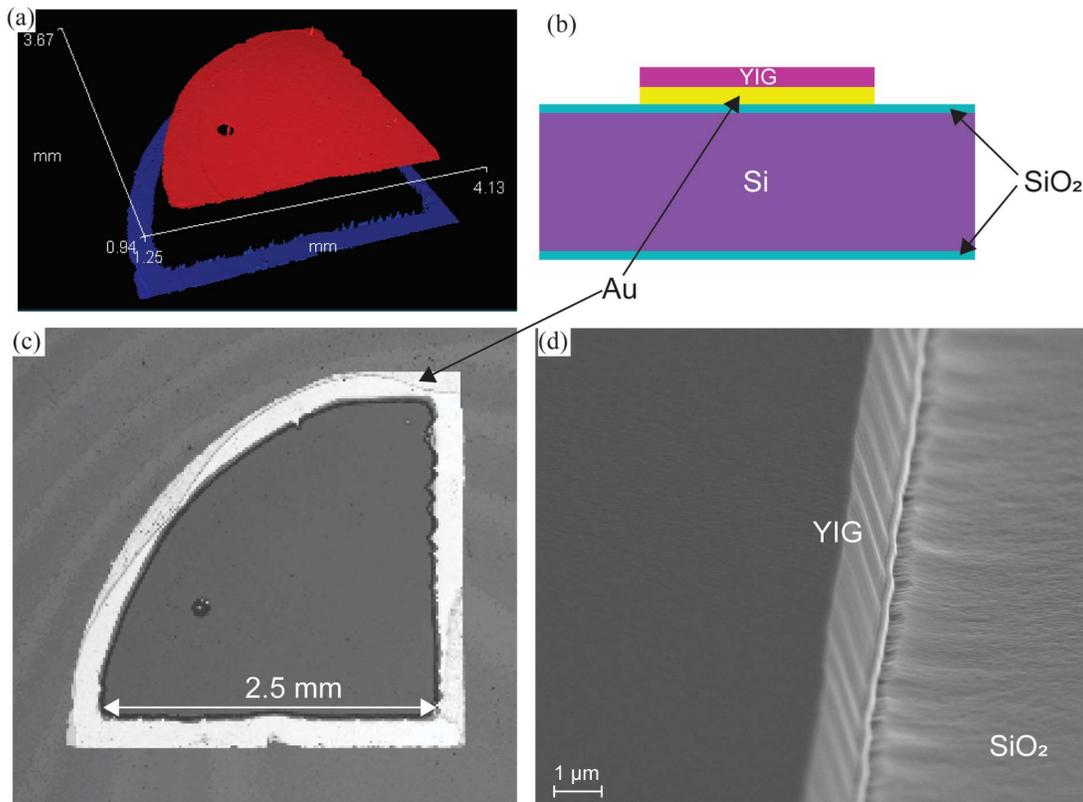

Figure 2: Ion-sliced and transferred YIG film on Si material platform: (a) laser interferometry measurement of ion-sliced YIG film on Si substrate; (b) cross-section schematic of the layer stack (from top to bottom: 2.19 µm YIG, 10 nm Ti, 60 nm Au, 10 nm Ti, 2 µm wet thermal oxide, 500 µm Si, 2 µm wet thermal oxide); (c) microscope photo of ion-sliced YIG film on Si, which was ion-sliced from a quarter of a 5mm diameter bulk single crystal YIG substrate; (d) SEM image of anisotropic etching profile of ion-mill patterning of YIG thin-film, etched through the gold bottom electrode stopping on the thermal oxide layer.

The measurement setup used for strain tunability is described in Figure 1. To strain the suspended YIG film in real-time, we actuate the silicon shuttles in the length direction by a linear translation stage (see Methods), and the actuation force is measured by a 6-axis load-cell. Meanwhile, a magnetic bias field from a 3D projection electromagnet is applied to the thin film. This allows us to experimentally characterize the frequency tuning by strain-induced anisotropy field of all three types of MSWs that can exist in the YIG thin-film. The magnetocrystalline anisotropy field of YIG in the (111) plane exhibits a 6-fold symmetry with respect to the crystal basis. However, it is relatively weak compared to the strain-induced anisotropy field relevant to this work. Therefore, without any loss of generality, we orient the length direction of our devices perpendicular to the crystal [1 -1 0] direction, which is determined by XRD before the first level of photolithography. On the other hand, the strain induced anisotropy exhibits a 2-fold symmetry with respect to the angle formed by the uniaxial stress and the saturation magnetization (which is always along the direction of applied DC magnetic bias). Therefore, we study the strain induced anisotropy field frequency tuning in three scenarios: when the static magnetic bias is aligned i) perpendicular and ii) parallel to the length direction of the device within the YIG thin-film plane, and iii) perpendicular to the film plane, while tensile straining the thin-film in the length direction.

When the static magnetic bias is in the thin-film plane and perpendicular to the length direction of the device, the resonator operates predominantly in the magnetostatic backward volume wave (MSBVW) mode due to the symmetry of the transducer (Figure 3), where the wave is excited through the out-of-plane component of the H field from the RF current flowing through the transducer fingers. The wave travels predominantly along the width direction of the suspended YIG film, parallel to the external applied magnetic bias. Figure 1 shows the measured impedance under a static magnetic bias of 526 Oe, while the suspended YIG thin film is strained in real-time by a linear translation stage from 0.06% to 1.06% of strain. Strong spurious modes exist on the lower frequency side of the main resonance. This is consistent with the fact that MSBVW exhibits anomalous dispersion. The measured resonant frequency of the main mode is tuned from 2.734 GHz to 4.571 GHz, resulting in a 1.837 GHz of tuning range with a tuning efficiency of 183.7 GHz/strain. This is consistent with our numerical model (see Methods). The 1.837 GHz of tuning is equivalent to an effective strain-induced magnetocrystalline anisotropy field of 642 Oe. Figure 3a also shows the impedance spectrum of a later iteration of the same design operating in the same configuration. This device is fabricated using an optimized process where the resputtered materials during ion-mill are removed by phosphoric acid at a temperature of 70°C for 25 minutes. This modified process produces a cleaner etching profile and results in sharper resonances in the spectrum. The measurement results shown in Figure 3a are with an in-plane bias of 1202 Oe. When no stress is applied, the resonant frequency is ~5.2 GHz. This is consistent with the theoretical value predicted for a 2.19 μm thick YIG film assuming a saturation magnetization of 1760 Oe (Methods), which indicates that the saturation magnetization of the suspended YIG film is consistent with the saturation magnetization of bulk single crystal YIG. When the uniaxial strain is varied from 0.1% to 0.28%, the resonant frequency is tuned from 5.35 GHz to 5.65 GHz, resulting in a tuning efficiency of 160.3 GHz/strain. As we further increase the external static magnetic bias, we notice a reduction in strain-induced frequency tuning efficiency as summarized in Figure 4. Intuitively, the total free energy at high bias is dominated by the Zeeman energy leading to a diminishing effect on the effective bias from change of magnetoelastic energy by straining.

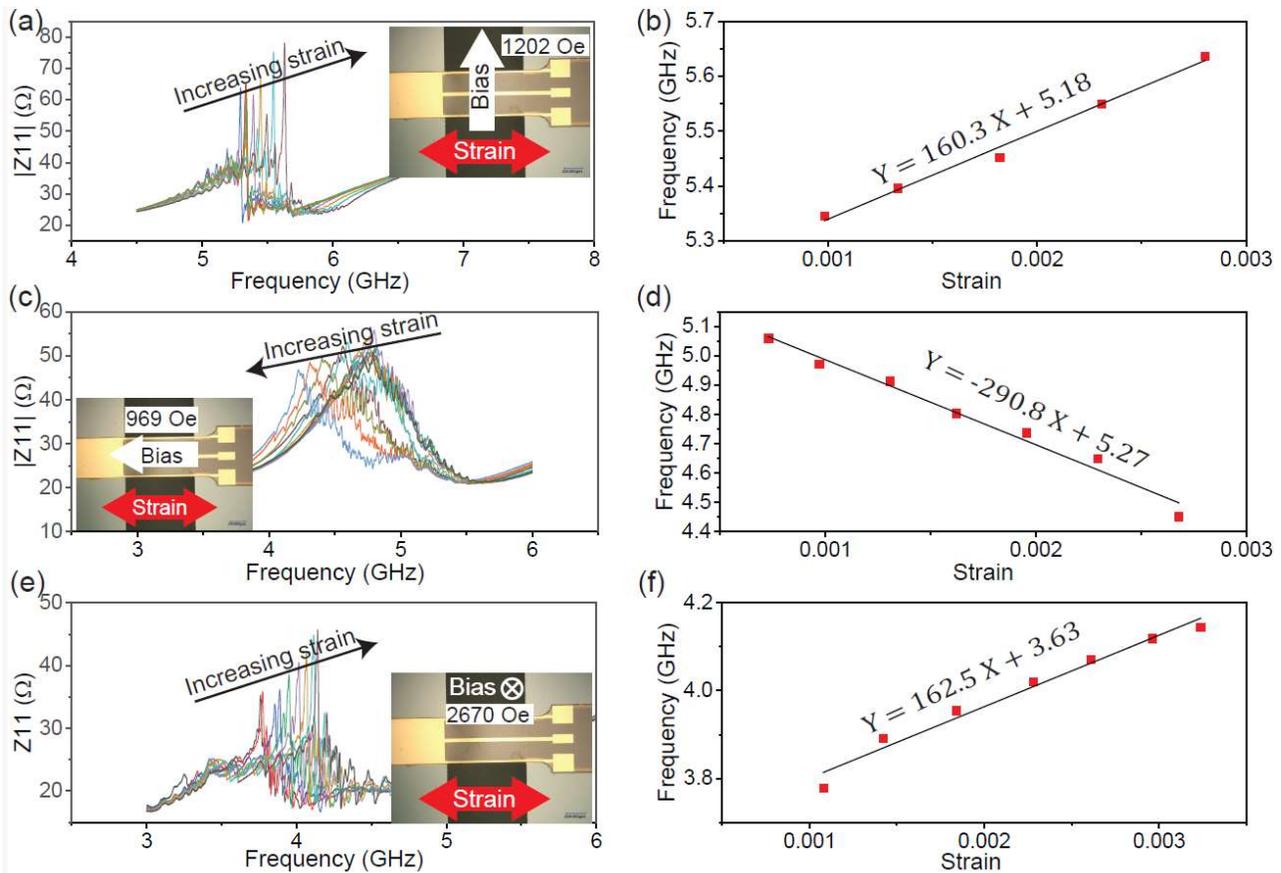

Figure 3: Measured strain-induced frequency tuning of suspended YIG thin-film MSW resonator operating in magnetostatic forward volume wave, magnetostatic backward volume wave, and magnetostatic surface wave configurations. Due to the anisotropic contribution of the magnetoelastic energy to the total free energy of magnetization, the frequency tuning efficiency for the three different configurations exhibits vast difference: (a) impedance of the thin-film YIG resonator operating in MSBVW configuration under strain-induced frequency tuning, inset annotates direction and magnitude of static magnetic bias and direction of strain; (b) resonant frequency vs. strain of the resonator operating in MSBVW configuration, showing a tuning efficiency of 160.3GHz/strain from linear regression fitting; (c) impedance of the same device operating in MSSW configuration under strain-induced frequency tuning, inset annotates direction and magnitude of static magnetic bias and direction of strain; (d) resonant frequency vs. strain of the resonator operating in MSSW configuration, showing a tuning efficiency of -290.8GHz/strain from linear regression fitting; (e) impedance of the same device operating in MSFVW configuration under strain-induced frequency tuning, inset annotates direction and magnitude of static magnetic bias and direction of strain; (f) resonant frequency vs. strain of the resonator operating in MSFVW configuration, showing a tuning efficiency of 162.5GHz/strain from linear regression fitting.

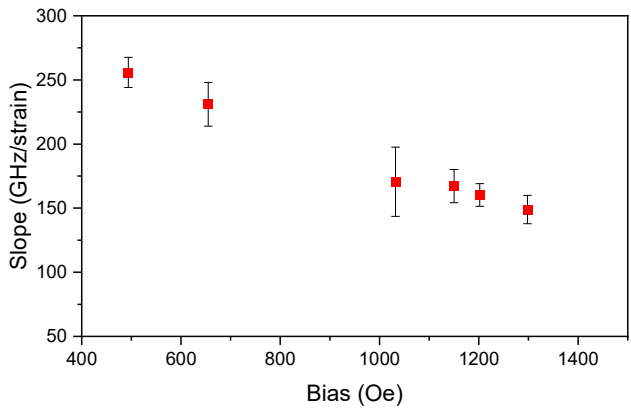

Figure 4: Strain tuning efficiency of MSBVW under different magnitudes of externally applied magnetic bias, showing decreasing tuning efficiency with increasing magnitude of externally applied static bias as the Zeeman energy dominates at high bias.

Figure 3c shows the impedance of the same device when a 969 Oe of external magnetic bias is applied parallel to the length direction of the thin film. In this configuration, the transducer predominantly couples to the magnetostatic surface waves (MSSW) through the in-plane component of the RF H-field. The wave travels predominantly along the width direction of the suspended thin-film, perpendicular to the externally applied magnetic bias. As surface wave exhibits normal dispersion, spurious models are predominantly on the higher frequency side of the main resonance. Consistent with our theoretical model, the resonant frequency of the device decreases as the suspended YIG is tensile strained, which is opposite to the MSBVW configuration. The tuning efficiency of MSSW is ~290.8 GHz/strain (Figure 3d), which is approximately 1.8 times higher than that of the MSBVW under a similar static magnetic bias. Finally, Figure 3e shows the measured strain-induced frequency tuning for the magnetostatic forward volume waves (MSFVW), where the magnetic bias is out of plane with a magnitude of 2670 Oe, and the wave propagates in the thin-film's width direction. In this configuration, the MSFVWs are excited through the in-plane component of the RF H-field from the transducer fingers. The tuning efficiency is 162.5 GHz/strain from linear fitting (Figure 3f). This is similar to that from the MSBVW configuration, and consistent with our numerical model.

Our results demonstrate a material and device platform that enables opportunities to explore abundant physics and engineering applications of the YIG material system. Combining a suspended YIG thin-film MSW resonator structure on a heterogeneous YoS material platform, we realized giant magneto-mechanical interactions in YIG, resulting in a record 1.837 GHz of real-time strain induced frequency tuning. Leveraging this device structure, we investigated the interaction of uniaxial tensile strain with the frequency tuning behavior of all three types of magnetostatic waves. An immediate advantage of this platform becomes apparent when we consider that the strain is a normalized quantity; hence, scaling the actuator dimension of the piezoelectric transducer appropriately can result in a much larger strain in the suspended YIG than the intrinsic strain capability of the piezoelectric material. In combination with our heterogeneous YIG-on-Si material platform, this would allow potential future integration of voltage-controlled, efficient, piezoelectric strain transduction using CMOS circuits.

*Acknowledgment*: The YIG on silicon substrate material platform was developed at FAST Labs[TM], BAE systems. Microfabrication including ion-milling of YIG, and back-side deep silicon etching was performed at the Birck Nanotechnology Center, Purdue University. Measurements were performed at Seng-Liang Wang Hall at Purdue University. The authors greatly appreciate the help and support from Prof. Michael Capano and Prof. Pavan Nukala on X-ray diffraction measurements. The views, opinions, and/or findings expressed are those of the authors and should not be interpreted as representing the official views or policies of the Department of Defense or the U.S. Government. This work was supported in part by the Air Force Research Laboratory (AFRL) and the Defense Advanced Research Projects Agency (DARPA).

**Authors Contributions**: R.W. invented device concept and design, completed modeling, and developed YoS material platform. Y.F. and S.D. performed short-loop fabrication runs to identify YOI release recipes and YIG material properties. S.T. worked closely with R.W. to update the design for testability and high yield, and micromachined the suspended YoS resonators. S.T. built measurement setup and conducted resonator measurements. R.W. and S.T. analyzed the experimental data. R.W. and S.T wrote the manuscript with input from others

†rw364@cornell.edu
*tiwari40@purdue.edu

# METHODS

## I. Magnetostatic wave dynamics

MSW are lattice waves, where the lattice consists of magnetic dipole precessions. Under a strong DC torque exerting bias, a localized RF disturbance sets the dipoles into precession, which then propagates through the material. Collectively, they form a propagating precession wave. In the classical regime (where wavelength is much longer than the exchange coupling length), the wave dynamics are governed by the Landau-Lifshitz-Gilbert (LLG) equation[49] coupled with Maxwell's equations under magneto-quasi-static approximation. The LLG equation governs the magnetic dipole precession dynamics, where $\boldsymbol{M}$ is the magnetic dipole moment per unit volume, and $\boldsymbol{H}_{eff}$ is the summation of any effective fields that can exert torque on the magnetic dipoles, The second term on the right-hand side of the equation accounts for the relaxation effect, suggested by T. L. Gilbert.

$$\frac{\partial \boldsymbol{M}}{\partial t} = -\gamma \boldsymbol{M} \times \boldsymbol{H}_{eff} + \frac{\alpha}{M_s} \boldsymbol{M} \times \frac{\partial \boldsymbol{M}}{\partial t}$$

As the wavelength approaches the characteristic exchange interaction length $\lambda_{ex}$, the exchange interaction among nearby electron spins become significant and gives rise to an extra torque exerting term ($\boldsymbol{H}_{ex} = \lambda_{ex}\nabla^2 \boldsymbol{M}$) in $\boldsymbol{H}_{eff}$. And magnetic dipole precession wave in such regime are referred to as spin waves. To facilitate the study of stress induced anisotropy field, we avoid the complexity arising from spin wave interactions by operating in the classical MSW regime in this work.

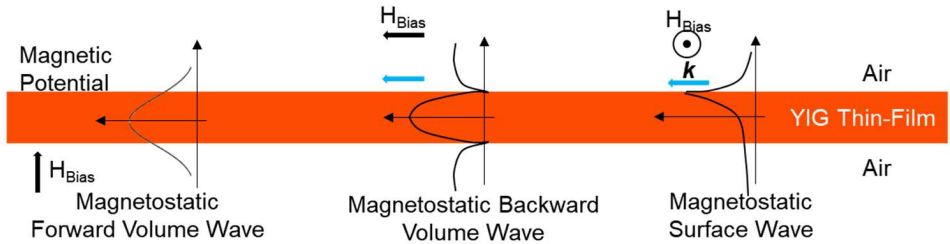

Figure 1. Types of magnetostatic waves that can be supported by thin-film ferro/ferromagnetic materials.

It is well known that a ferro/ferromagnetic thin film structure, in general, supports three types of MSW due to the 2D confinement of the film. Namely, there are magnetostatic forward volume waves, magnetostatic backward volume waves, and magnetostatic surface waves. The particular type of waves that can be supported and excited in a thin-film structure depends on the configuration of the directions of the static magnetic bias, the wave-vector, and the RF excitation field, which are annotated in Figure 2. Our MSW transducer is designed to couple to all wave types depending on the static bias configuration to facilitate the study of stress induced anisotropy field tuning, albeit the coupling efficiencies are different for different types of waves.

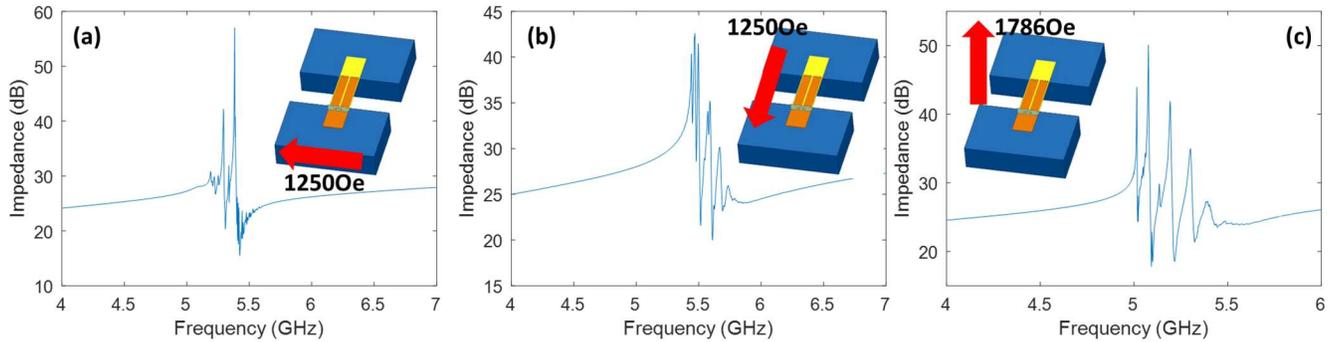

Figure 2: HFSS simulation of small signal RF response (impedance) of the suspended YIG resonator devices operating in the MSBVW, MSSW, and MSFVW configurations: (a) simulated impedance of the resonator operating in the MSBVW configuration under 1250Oe static magnetic bias; (b) simulated impedance of the resonator operating in the MSSW configuration under 1250Oe static magnetic bias; (c) simulated impedance of the resonator operating in the MSFVW configuration under 1786Oe out of plan static magnetic bias.

In addition, it has been shown that the dispersion relations for the lowest order MSWs can be written as[50]

$$\text{MSFVW:} \quad \omega^2 = \omega_0 \left[\omega_0 + \omega_M \left(1 - \frac{1-e^{-kd}}{kd}\right)\right] \tag{1}$$

$$\text{MSBVW:} \quad \omega^2 = \omega_0 \left[\omega_0 + \omega_M \left(\frac{1-e^{-kd}}{kd}\right)\right] \tag{2}$$

$$\text{MSSW:} \quad \omega^2 = \omega_0(\omega_0 + \omega_M) + \frac{\omega_M^2}{4}(1 - e^{-2kd}) \tag{3}$$

where $\omega_0 = \mu_0 \gamma H_{eff}^{DC}$, $\omega_M = \mu_0 \gamma M_s$, and $\omega$ and $k$ are respectively the frequency and wave-vector of the MSFW (parallel to film), $d$ is the film thickness, $\gamma$ is the gyromagnetic ratio, $\mu_0$ is the vacuum permeability, $M_s$ is the magnitude of the saturation magnetization. For our devices, we operate in the regime where $kd \ll 1$ (thin-film approximation), so that the tuning efficiency of MSW resonances approaches that of uniform precession mode. We model the small signal RF behavior of our devices using ANSYS HFSS, as the physics is captured by Maxwell's equations and the permeability tensor from linearized LLG equation. Figure 2a shows the simulated resonator impedance as a function of frequency for the device in main manuscript operating in the MSBVW configuration. The internal magnetic bias is 1250Oe in the plane of the YIG thin film and perpendicular to the device length direction. Spurious modes exists on the lower frequency side of the main resonance, consistent with our measurements. Figure 2b shows the simulated impedance of the same device operating in the MSSW configuration with an internal magnetic bias of 1250Oe, while Figure 2c shows the impedance of the device operating in the MSFVW wave with an out-of-plane internal magnetic bias of 1786Oe.

## II. Tuning of Resonator Frequency through Strain Induced Anisotropy Field

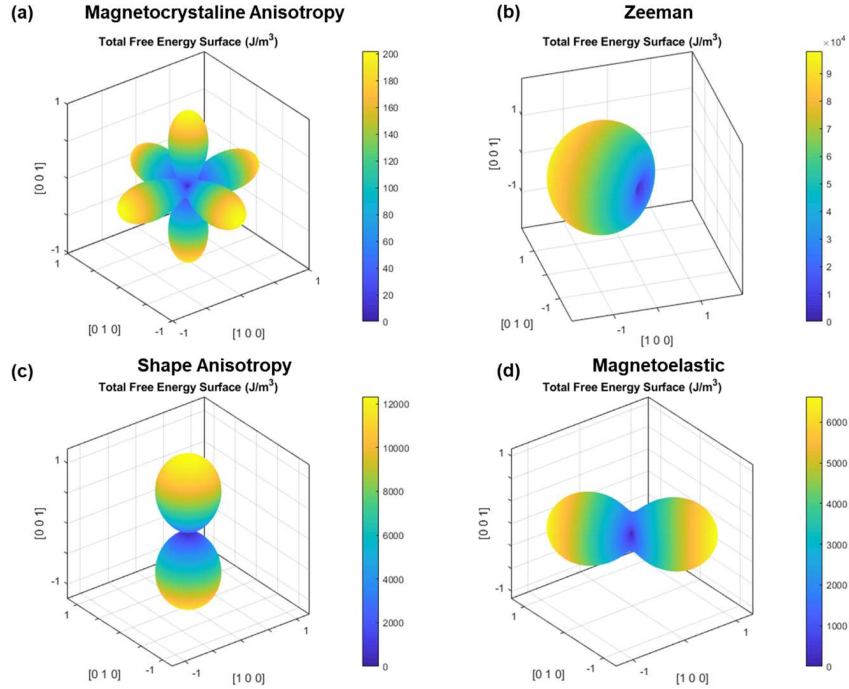

Figure 3: Decomposition of the total free energy into its individual components: (a) total free energy surface only due to the magnetocrystalline anisotropy; (b) total free energy surface only due to the Zeeman energy; (c) total free energy surface only due to the demagnetization field; (d) total free energy surface only due to the magnetoelastic effect.

It has been shown that the uniform magnetic dipole precession frequency of a ferromagnetic body, hence the effective magnetic bias, can be calculated as[51]

$$\omega_0 = \mu_0 \gamma H_{eff}^{DC} = \frac{\gamma}{M_s \sin\theta} \sqrt{\frac{\partial^2 E}{\partial \theta^2}\frac{\partial^2 E}{\partial \varphi^2} - \left(\frac{\partial^2 E}{\partial \theta \partial \varphi}\right)^2} \quad (4)$$

where $\theta$ and $\varphi$ are the polar coordinates, $\gamma$ is the gyromagnetic ratio, $\mu_0$ is the vacuum permeability, $H_{eff}^{DC}$ is the effective magnetic bias, and $E$ is the total free energy of the magnetic dipole moments. Many fields that can exert torque on the dipole moments contribute to the total free energy. As it is assumed that a strong static bias is always present during the operation of our devices, and the wavelength considered here is much longer than the exchange-coupling length in YIG[52], we do not consider the exchange field in our formulation. Here, we consider the energy from static magnetic bias (Zeeman energy[1]), demagnetization field[53], magnetocrystalline anisotropy field[54], and stress induced magnetocrystalline anisotropy field (magnetoelastic energy[42,55]), and they can be written as below

$$E_{Zeeman} = -\mu_0 M_S \cdot H_{Ext} \quad (5)$$
$$E_{Demag} = \mu_0 M_S \cdot H_{Demag}/2 \quad (6)$$
$$E_{Ani} = K_0 + K_1(\alpha_1^2\alpha_2^2 + \alpha_2^2\alpha_3^2 + \alpha_3^2\alpha_1^2) + K_2(\alpha_1^2\alpha_2^2\alpha_3^2) + \cdots \quad (7)$$
$$E_{ME} = -3\lambda_{000}(\alpha_1^2\sigma_{11} + \alpha_2^2\sigma_{22} + \alpha_3^2\sigma_{33})/2 - 3\lambda_{111}(\alpha_1\alpha_2\sigma_{12} + \alpha_2\alpha_3\sigma_{23} + \alpha_3\alpha_1\sigma_{31}) \quad (8)$$

where $M_s$ is the saturation magnetization, $H_{Ext}$ is the externally applied static magnetic bias, $H_{Demag}$ is the demagnetization field (shape anisotropy field), $K_i$ are the crystalline anisotropy constants, $\alpha_i$ are the directional consines of the saturation magnetization with respect to the crystal axis, $\lambda_{ijk}$ are the magnetostriction constants, and $\sigma_{ij}$ are the components of the stress tensor. To model the strain-induced anisotropy field, we start with calculating the total free energy. For example, Figure 4b shows the total free energy surface of a (001) YIG thin film, with a 3500Oe external static magnetic bias and a 1% uniaxial strain both applied in the [1 -1 0] direction, where the magnitude of the vector connecting the origin to the points on the surface denotes the total free energy. The free energy surface exhibits a global minimum in the [1 -1 0] direction, which is consistent with fact that the strong static bias dominates that potential energy so that the saturation magnetization is aligned in [1 -1 0]. Figure 3 shows the decomposition of the free energies into their individual components caused by only the magnetocrystalline anisotropy energy (Figure 3a), Zeeman energy (Figure 3b), demagnetization energy (Figure 3c), and strain-induced crystalline anisotropy energy (Figure 3d). As anticipated, the free energy surface, due only to the magnetocrystalline anisotropy exhibits six maximum energy lobes pointing to the [001] equivalent directions, and 8 minimum energy points in the [111] equivalent directions. This is consistent with the fact that the [001] directions are the hard axis of YIG while [111] are the soft axis of YIG. Meanwhile, the free energy surface due only to the demagnetization field are minimized in the (001) plane, consistent with that the shape anisotropy of a ferromagnetic thin film tends to align the magnetization within the thin film. The Zeeman energy surface exhibits a global minimum at the [1 -1 0] direction, which is consistent with that the external applied magnetic bias is in the [1 -1 0] direction. Using eq. 4, we then numerically calculate the uniform precession resonant frequency (Figure 4d) and the effective magnetic bias, which results in a resonant frequency of 9.66GHz or equivalently an effective bias of 2681Oe calculated using Kittel's formula[53] assuming a saturation magnetization of 1760 Oe for YIG. It is worth noting that the resonant frequency surface only has physical meaning in the direction of the global minimum of the free energy surface. Comparing to the exact same configuration without any strain in YIG (Figure 4a, c), the resonant frequency is 12.11GHz or equivalently an effective bias of 3534Oe. The minor difference in the effective bias calculated from Kittel's formula versus the 3500Oe model input was due to the effect of crystalline anisotropy of YIG.

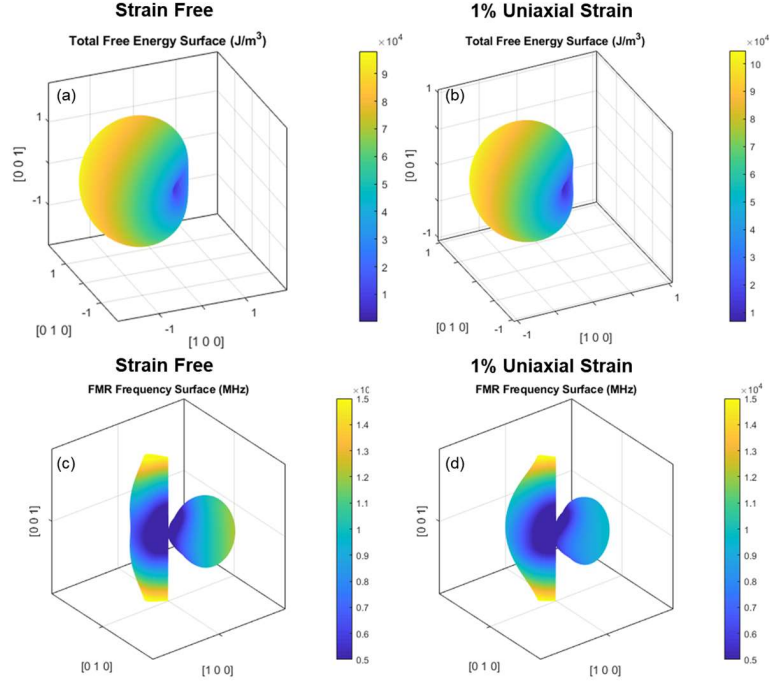

Figure 4: Numerical calculation of effective bias from strain-induced anisotropy field: (a) total free energy when the material is stress free; (b) total free energy when the thin-film is subjected to 1% of uniaxial strain in the [1 -1 0] direction; (c) calculated frequency surface from the total free energy surface in (a); calculated frequency surface from the total free energy surface in (b).

As indicated by eq. (5) ~ (8), the effective bias depends on the direction of the $M_s$ as well as the stress tensor with respect to the crystal axis. In Figure 5, we calculate the stress-induced frequency change of a (111) YIG thin-film under 2000 Oe of in-plane static magnetic bias, when an in-plane uniaxial stress of 1% is applied. The simulation is performed for different directions of uniaxial stress with respect to the crystal basis and for different directions of static magnetic bias with respect to the direction of the uniaxial stress. As shown in the figure, the strain-induced frequency tuning exhibits a 2-fold symmetry with respect to the direction of the bias. When the bias is parallel to the uniaxial stress, the frequency tuning is -2.35GHz at 1% strain comparing to no strain. Meanwhile, when the bias is perpendicular to the uniaxial stress, the frequency tuning is positive 1.2GHz. In addition, the frequency change shows a very weak 6-fold symmetry with respect to the direction of the uniaxial stress, which is due to the weak interaction between the strain-induced anisotropy and the magnetocrystalline anisotropy.

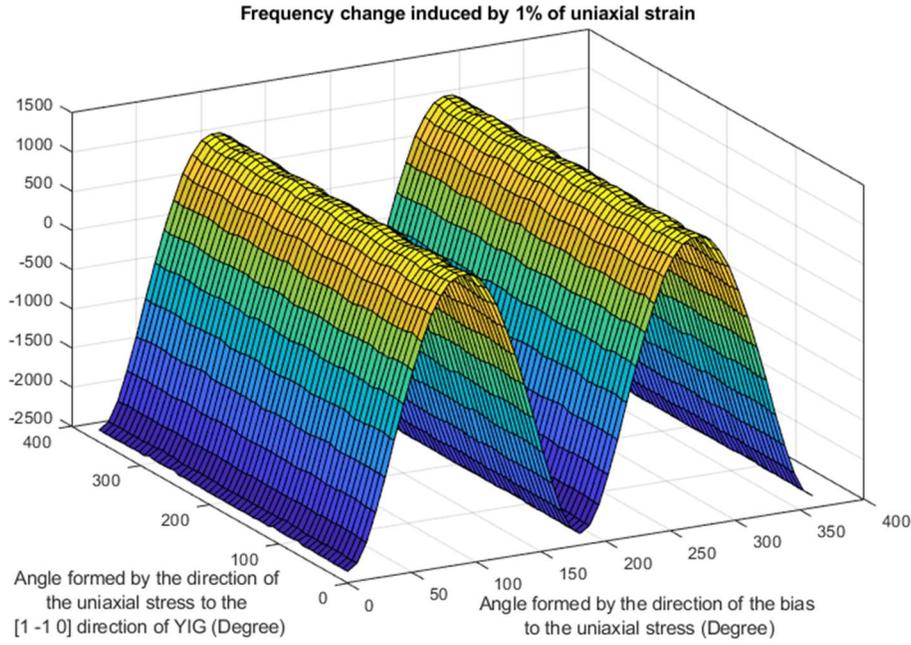

Figure 5: Strain-induced frequency tuning in a (111) YIG thin film under an in-plane uniaxial strain of 1% and an in-plane magnetic bias of 2000Oe.

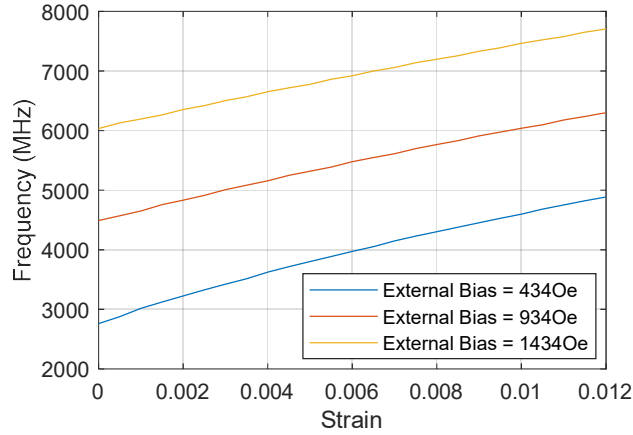

Figure 6: Strain-induced frequency tuning for MSBVW under different externally applied magnetic bias.

Figure 6 shows the calculated strain-induced frequency tuning under different externally applied magnetic bias. Here, the uniaxial stress is parallel to the YIG [1 -1 0] direction and the static bias is perpendicular to the [1 -1 0] direction, which corresponds to the MSBVW case in our measurements. At an external bias of 434Oe, the frequency change at 1.1% strain is 1.95GHz. This is consistent with our measurement in Figure 1, where a 1.02% strain led to 1.837GHz of frequency tuning. The frequency tuning efficiency from straining exhibits weak dependence on the magnitude of externally applied magnetic bias. The tuning efficiency at 434Oe, 934Oe, and 1434Oe are 177GHz/strain, 151GHz/strain, and 139GHz/strain, respectively. In addition, Figure 7 shows the frequency tuning as a function of strain for different wave configuration under different externally applied magnetic bias. The orange curve shows tuning corresponding to the MSBVW case, similar to Figure 6 but at an external bias of 1202Oe. The tuning efficiency is 148GHz/strain. Meanwhile, the yellow curve is for the case where the uniaxial stress is parallel to the YIG [1 -1 0] direction and the static bias is also parallel to the [1 -1 0] direction with a static bias of 969Oe, corresponding to the MSSW case in our measurements. The tuning efficiency is 294GHz/strain, which matches with our measurements well. The simulated tuning efficiency of MSSW configuration at this bias is about 2 times of that of the MSBVW case, while the measured tuning efficiency for the MSSW case is about 1.8 times of the MSBVW. In addition, the blue curve is for the case where the uniaxial stress is parallel to the YIG [1 -1 0] direction and the static bias is normal to the thin-film with a static bias of 2670Oe, corresponding to the MSFVW case in our measurements. The tuning efficiency is about 12% lower than that of the MSBVW case in the model. In comparison, our measurements indicated that the tuning efficiencies are similar. We attribute these numerical mismatches between model and measurements to the inaccuracy of the material parameters used in the model as well as the fact that the stress in the suspended YIG thin-film is not strictly uniaxial.

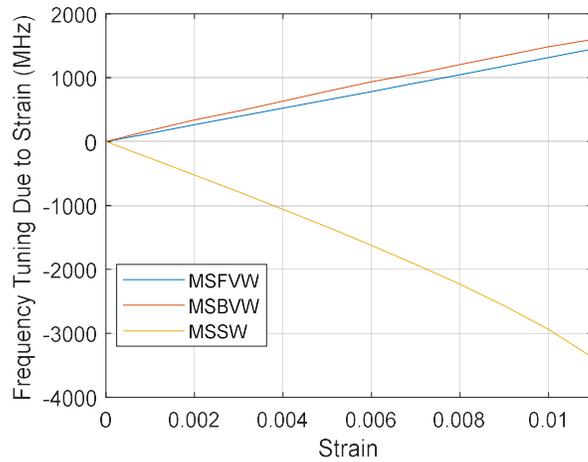
Figure 7: Strain-induced frequency tuning as a function of strain induced by uniaxial stress for static bias perpendicular and parallel to the uniaxial stress.

## III. Fabrication process flow

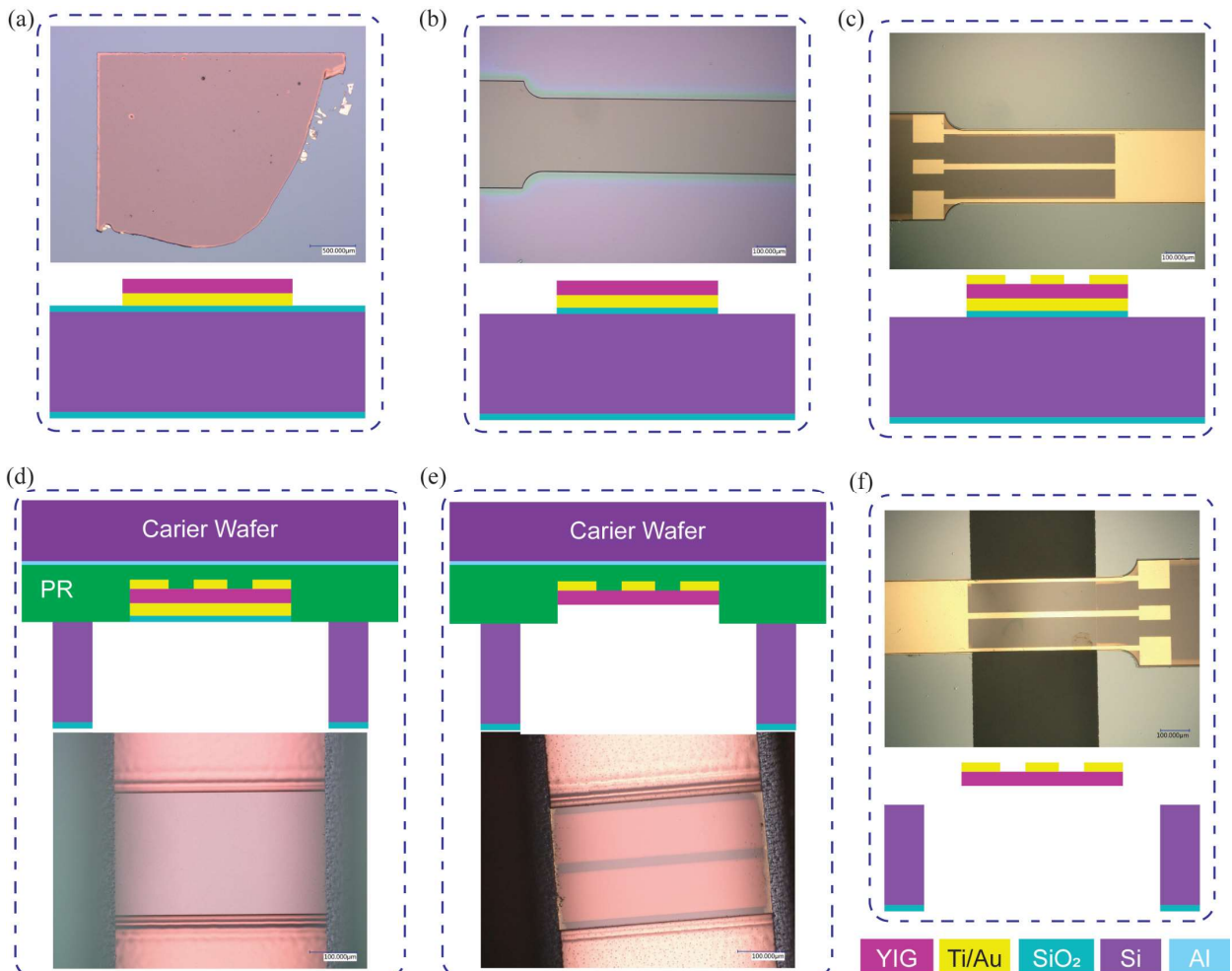
Figure 8: Microfabrication process flow of the suspended YIG thin-film devices. The (a) Starting YoS material stack, (b) Optical image and corresponding cross-sectional schematic after patterning of YIG layer, (c) Optical image and corresponding cross-sectional schematic after patterned deposition of Ti/Au layers which serve as transducers. (d) Cross-sectional schematic and an optical image of the sample after backside DRIE of Si layer. (e) Cross-sectional schematic and an optical image of the sample after etching of SiO₂, and bonding metallic layers. The sample is on a carier wafer for steps (d-e). (f) Optical image and a scross-sectional schematic after demounting of the sample from the carier wafer.

The fabrication process (Figure 8) begins with a thin-film YoS substrate (Main manuscript), which is fabricated by an ion-slicing, and thin-film transferred process. During the ion-slicing process, a bulk single crystal YIG grown by floating zone method is irradiated with 1MeV He$^+$ ions. This creates a damaged layer centered at ~3 μm blow the YIG surface. Next, the YIG is flipped bonded on a high resistivity Si wafer with 2 μm thick thermal oxides on both sides by gold to gold compression bonding process. The bonding layer consists of 10nm of Ti adhesion layer and 30nm of Au bonding layer on both the Si and the YIG. This is followed by an anneal process that slice off the YIG film from the plane damaged by the high energy helium ions. A phosphoric acid etch at 65C with 85% concentration selectively removes the ion-implantation damaged surface layer on the sliced-off YIG, which results in a YIG thin-film of around 2.4 μm thickness (Figure 8a). The high Si substrate resistivity minimize RF loss for the final resonator device. As the bonding strength of YIG on the Si wafer is crucial for the straining of the suspended YIG thin-film, the bonding strength has been tested by die shearing test indicating a >100MPa bonding shear strength. In addition, samples have been thermally cycled to over 450°C, and IR imaging of the bonding interface showed no signs of delamination. As the thermal expansion coefficients of YIG and Si are 10.4ppm/°C and 2.6ppm/°C, the peak shear thermal stress at the bonding interface from the thermal cycling exceeds 300MPa, indicating extremely high bonding quality. The microfabrication process flow after on the ion sliced YIG-on-Si wafer is shows in Fig. 8. The cross-sectional schematics are accompanied by an optical image of the sample at each step. The optical from topside of the wafer are situated over the cross-sectional schematic while the bottom side optical images are situated under the schematic. After slicing off, the YIG films (>2 μm) are patterned through ion milling utilizing an AJA International, Inc. system. An optimizes photoresist mask is used to avoid burning of photo-resist during this thick YIG etching (Figure 8b). This ion milling process causes some sputtering of etched material on the vertical sidewalls. To remove this resputtered material, the sample is immersed in phosphoric acid at a temperature of 70 °C for 25 minutes. Consequently, this combined etching procedure results in an almost vertical sidewall profile. Additionally, the phosphoric acid soak reduces the thickness of the YIG film by 100 nm. During the ion milling process for YIG etching, an over-etch is carried out to etch the Au/Ti layers, which were used for bonding the YIG to the Si wafer. Consequently, the buried SiO$_2$ layer is exposed. This 2 μm SiO$_2$ layer is removed using reactive ion etching (RIE). Next, a top electrode consisting of 10nm Ti adhesion layer and 300nm of Au is defined over the YIG by lift-off (Figure 8c) process, which serves as the magnetostatic wave transducer. After the front-side process, the YoS substrate is flip-mounted on a carrier wafer for back-side processing. To protect the top side of the sample, a photoresist layer is spin coated, and a 100 nm layer of Aluminum (Al) is deposited. The Al layer serves to prevent the formation of a difficult-to-clean mixture of photoresist and crystal-bond 555, which is utilized for bonding the sample to a carrier wafer. Next, deep reactive ion etching (DRIE) is used to etch through the bulk of the Si wafer. The thermal oxide on the backside of the Si wafer is patterned by RIE using photoresist masks, which serves as a hardmask for the DRIE process. The buried thermal oxide between the bonding metal and Si serves as the etch stop for the DRIE (Figure 8d), which is exposed after the DRIE etch. The buried SiO$_2$ layer is then partially etched using reactive ion etching. The remaining SiO$_2$ layer and bonding metal layers (Ti/Au/Ti) are removed using wet chemical processes (Figure 8e). The removal of the gold layer is accomplished using a standard Au etchant, while BOE is employed for the removal of the SiO$_2$ and Ti layers. Throughout this entire process, the sample remains attached to the carrier wafer. Subsequently, the sample is demounted by immersing the carrier wafer in hot water. Following the demounting, the sample is soaked in acetone to lift off the protective Al layer. As a result of this process, a suspended YIG film with MSW transducers on top is obtained (Figure 8f).

## IV. Measurements

The strain tuning measurement requires force-displacement measurement simultaneously with RF measurement under a magnetic bias field. In order to achieve this feature, a custom test setup was designed and assembled. The measurement setup consists of a linear translation stage (X-LDM060C from Zaber Technologies Inc) to provide necessary movements, thereby enabling strain of YIG film, a 6-axis load cell (MC3A from Advanced Mechanical Technology, Inc.) to completely characterize the forces and moments generated during the movement of the linear stage, a projection magnet (Model 5201 from GMW Associates) and a typical RF measurement setup (i.e., VNA and GSG RF probes along with a microscope). Two machined aluminum plates, one each, are mounted on the linear stage and the load cell. These plates are machined with a M1.4 screw whole where a partially screwed M1.4 screw is fixed. The sample is mounted between the linear stage and load cell by aligning the holes in the Si with the screws. This allows the stretching of the Si plates, which in turn transfers stresses to the YIG thin film. The measurement setup is schematically shown in Figure 9. The equivalent spring model of this measurement is shown in Figure 10. Both the load cell and the linear stage have their own stiffnesses as denoted by $K_{LC}$ and $K_{LS}$, which are connected in series with the device where the suspended YIG film and the straight Si springs are in parallel combination. When the linear stage is actuated, the complete spring assembly comes under tension, and a fraction of the total force is transferred to the YIG film. After each stage movement the force reading is allowed to stabilize, and the corresponding RF response of the device is saved using the VNA. A constant bias field is maintained using the projection magnet, and measurements are taken for incremental displacement intervals. This process is repeated for a combination of bias fields in all three axes. The spring model can be further simplified by using an effective spring of stiffness, $K_C$ which, accounts for $K_{LS}$ and $K_{LC}$ in series. To measure this stage assembly stiffness $K_C$, a rigid Si block with the same mounting hole but without any spring is fabricated. A force vs. displacement response is measured after mounting this Si block on the measurement setup. Since the Si block can be assumed rigid for this measurement, the slope of this curve yields the stiffness of the stage assembly. For accurate characterization of the spring constant of the Si springs, the force displacement measurement is carried out on a device with ruptured YIG film. From this measurement, the combined stiffness of stage assembly and Si springs ($K_{Si}$ and $K_C$ in series) is obtained. Since $K_C$ has already been calculated, $K_{Si}$ can be easily calculated. From the different stiffness measurements the fraction of total force that is experienced by the YIG film is calculated. This force is used to calculate the strain in the YIG film for each measurement point.

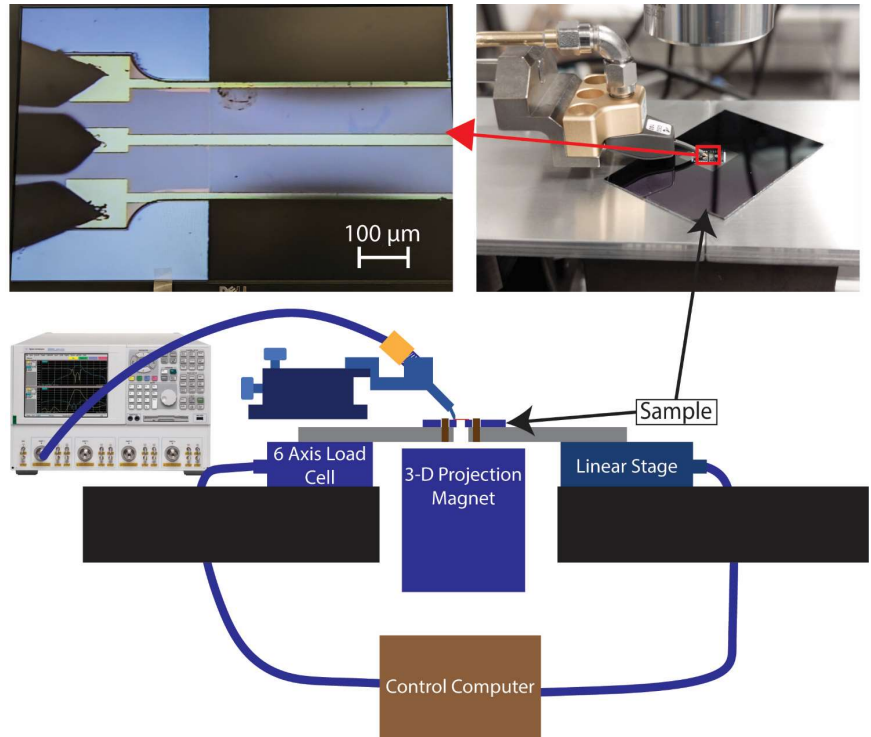

Figure 9: Setup for testing the suspended thin-film YIG devices.

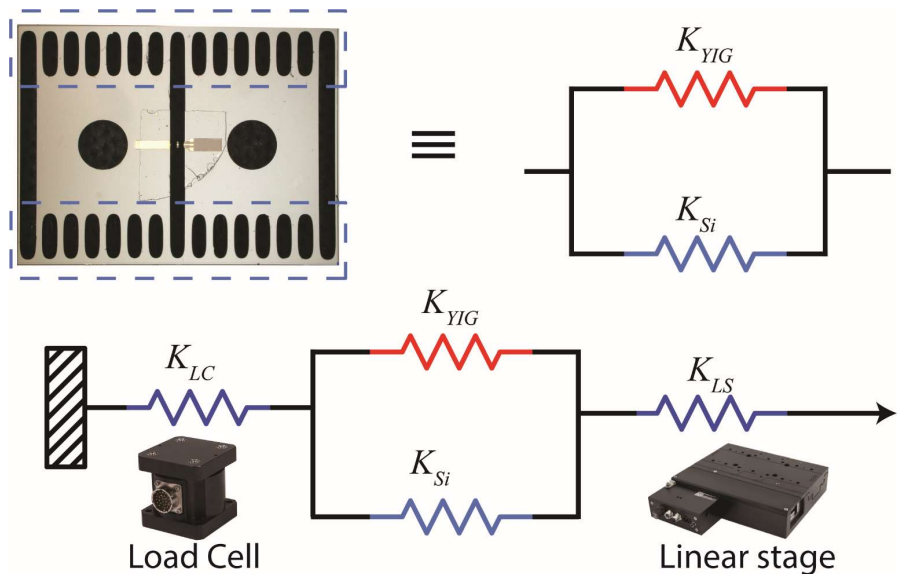

Figure 10: The equivalent spring connection model of the full characterization setup.